\documentclass[aps,prl,twocolumn,superscriptaddress,floatfix,longbibliography]{revtex4-2}

\usepackage{amsmath}
\usepackage{amssymb}
\usepackage{graphicx}
\usepackage{bm}
\usepackage[colorlinks=true,allcolors=blue]{hyperref}

\usepackage{xcolor}
\usepackage{graphicx}
\usepackage{dcolumn}
\usepackage{bm}
\usepackage{tikz}
\usepackage[caption=false]{subfig}

\newcommand{\half}{\frac{1}{2}}

\graphicspath{{figures/}}

\newcommand{\M}{\mathbf{M}}
\newcommand{\dm}{\delta\mathbf{m}}

\begin{document}

\title{Nonlinear Antimagnonics}

\author{A.~L.~Bassant}
\email{a.l.bassant@uu.nl.}
\affiliation{Institute for Theoretical Physics, Utrecht University, Princetonplein 5, 3584 CC Utrecht, The Netherlands}

\author{R.~A.~Duine}
\affiliation{Institute for Theoretical Physics, Utrecht University, Princetonplein 5, 3584 CC Utrecht, The Netherlands}
\affiliation{Department of Applied Physics, Eindhoven University of Technology, P.O. Box 513, 5600 MB Eindhoven, The Netherlands}

\author{G.~E.~W.~Bauer}
\affiliation{Advanced Institute for Materials Research, Tohoku University, Sendai 980-8577, Japan}
 \affiliation{Center for Science and Innovation in Spintronics, Tohoku University, 2-1-1 Katahira, Sendai 980-8577, Japan}

\author{M.~Elyasi}
\affiliation{Advanced Institute for Materials Research, Tohoku University, Sendai 980-8577, Japan}
 \affiliation{Center for Science and Innovation in Spintronics, Tohoku University, 2-1-1 Katahira, Sendai 980-8577, Japan}

\date{\today}

\begin{abstract}
The dynamics of ferromagnets are a resource of nonlinearities that are essential for quantum signal processing and state control, but to date the required microwave driving fields appear to be too large.
Here we predict substantially enhanced nonlinearities such as a reduced threshold for the ``foldover" of an anharmonic oscillator in the inverted magnetization state under current-induced spin transfer torque in isotropic magnetic films. The dipolar field of the precessing inverted macrospin parametrically excites ``antimagnons" and associated phase-locking and frequency combs at unprecedentedly weak microwave powers.
Our findings establish the inverted magnetization state as a superior platform for spintronic signal processing, sensing, and quantum (anti)magnonics.
\end{abstract}

\maketitle

\textit{Introduction.}---The dynamic stabilization of an \emph{inverted} magnetization, i.e., opposite to the equilibrium direction, can be achieved by current-induced spin-transfer torque (STT) in magnetic tunnel junctions \cite{Sun_2000-7-1,Grollier_Cros_Jaffrès_Hamzic_George_Faini_Ben_Youssef_Le_Gall_Fert_2003-5-5,Ozyilmaz_Kent_Monsma_Sun_Rooks_Koch_2003} and in bilayers of a magnet and a heavy metal ~\cite{Kurebayashi_Barker_Yamazaki_Kushwaha_Stenning_Youel_Hou_Dion_Prestwood_Bauer_etal._2026,Karadza_Wang_Kercher_Noel_Legrand_Schlitz_Gambardella_2026}.
Its dynamics unlock functionalities required for stochastic computing~\cite{Kurebayashi_Barker_Yamazaki_Kushwaha_Stenning_Youel_Hou_Dion_Prestwood_Bauer_etal._2026}.
Here we address the potential of wave-like excitations in the inverted state, quantized as \emph{antimagnons}, which have recently enriched the field of magnonics~\cite{AntimagnonEXP, Harms_Yuan_Duine_2024}.
Antimagnons emulate field-theoretic effects such as bosonic Klein scattering~\cite{Harms,Adorno2024,bassant2025phonon} and magnon-antimagnon pair production~\cite{Bassant,Kleinherbers}, amplify magnon currents, and generate entangled states.

\begin{figure}[t]
\includegraphics[width=\linewidth]{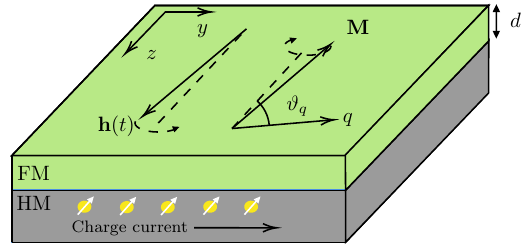}
\caption{\label{fig:setup}%
The magnetization $\mathbf{M}$ of a thin ferromagnetic film (FM) of thickness $d$  is subject to a magnetic field $\mathbf{h}(t)=h_z\mathbf{e}_z+\mathbf{h}_\perp(t)$. The charge current and associated spin Hall effect in an attached heavy metal (HM) layer generate a transverse spin current and a spin-transfer torque (STT) that, above a critical value, inverts the magnetization opposite to \(h_z\mathbf{e}_z\).
Its excitations are antimagnons with wave number $q$ and in-plane propagation angle $\vartheta_q$.
The microwave field \(\mathbf{h}_\perp(t)\) induces a precession around $-\mathbf{e}_z$ and dipolar fields that excite antimagnon frequency combs and plateaus.}
\end{figure}

To date, antimagnons have been treated with a linearized theory, missing the nonlinearities that are central to magnonics and enable advanced computing and signal-processing functionalities.
Nonlinear phenomena have been demonstrated in magnetic systems, while the devices they enable remain largely optical or electronic: multistability~\cite{Duf:Moon2014,Multistable:PhysRevLett.127.183202} stores information in coexisting states, solitons~\cite{KOSEVICH1990117} transmit signals without distortion~\cite{Hasegawa_2022,Mollenauer_Smith_1988}, frequency combs~\cite{Carmiggelt2023,NSW:10.1063/5.0090033,NSW:Koerner_Dreyer_Wagener_Liebing_Bauer_Woltersdorf_2022} are used for frequency conversion in metrology and telecommunications~\cite{Fortier_Baumann_2019}, and synchronization~\cite{NSW:Bauer2,Synchronisation:PhysRevB.90.054414,STTdyn:RobertoBonin_2010,STTdyn:10.1063/1.2712324} phase-locks oscillators into coherent sources~\cite{Wetzel_Jörg_Pollakis_Rave_Fettweis_Jülicher_2017}.
Nonlinearity is also indispensable for quantum magnonics, since linear theories and harmonic oscillators cannot generate the non-Gaussian distributions required for entanglement-based protocols~\cite{QM:Hioki2026,QM:PhysRevB.101.054402}.

A central issue in quantum magnonics is the too-high power threshold of nonlinear processes. In this Letter, we show that nonlinearities come by at much lower microwave powers at the cost of having to apply a current that can be minimized by a nearly vanishing magnetic anisotropy \cite{Kurebayashi_Barker_Yamazaki_Kushwaha_Stenning_Youel_Hou_Dion_Prestwood_Bauer_etal._2026,Karadza_Wang_Kercher_Noel_Legrand_Schlitz_Gambardella_2026}.

We consider a thin ferromagnetic (FM) film on a heavy metal (HM) layer that supplies STT via the spin Hall effect, as sketched in Fig.~\ref{fig:setup}.
The FM is subject to a static field $h_z$ and to an oscillating transverse field of amplitude $h_\perp$ at frequency $\omega$ that, at resonance, drives the magnetization into a steady precession.
The dynamic dipolar fields associated with uniform precession near the equilibrium and inverted uniform magnetization parametrically modulate non-uniform spin waves in both configurations. Here, we show that this modulation is remarkably more efficient in the latter case, generating resonance and frequency combs even at relatively weak microwave power, which facilitates signal processing and sensing applications.

\begin{figure}
    \centering

    {
        \includegraphics[width=\linewidth]{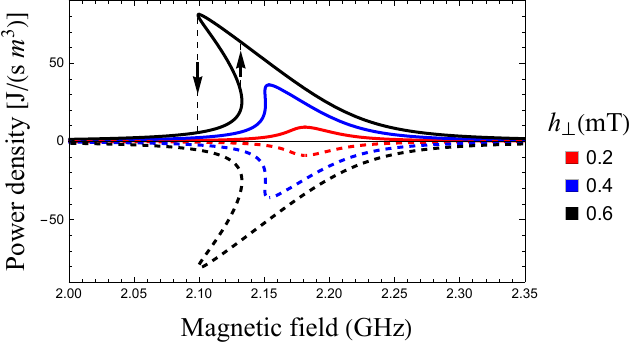}
        \label{fig:FMRhp}
    }

    {
        \includegraphics[width=\linewidth]{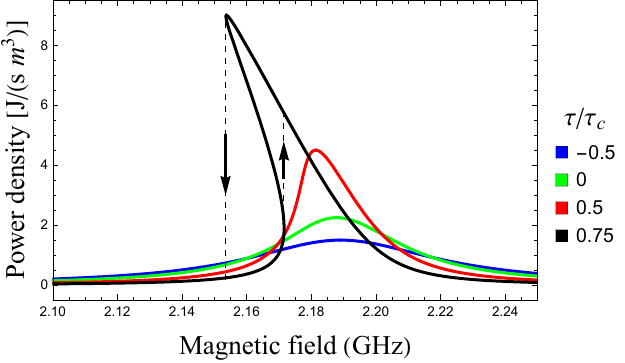}
        \label{fig:FMRt}
    }

    \caption{Absorption power of microwaves $p$ at a fixed frequency $\omega=1.89~\mathrm{GHz}$ as a function of the static magnetic field $h_z$  with and without current-induced STT $\tau$.
(a) For different microwave amplitudes $h_\perp$ at zero STT (solid) and at twice the critical STT ($\tau=2\tau_c$) for inversion (dashed).
(b) Several $\tau$ at fixed $h_\perp=0.1\text{ mT}$.
Both panels show fold-overs produced by the anisotropy, which renders the response hysteretic in $h_z$: sweeping up follows the upward arrow, sweeping down the downward one.
Parameters: $M_s=100~\mathrm{kA/m}$, $\gamma=28.134~\mathrm{GHz/T}$, $\alpha=0.0125$, $\omega=1.89~\mathrm{GHz}$, and $\gamma\kappa=0.3~\mathrm{GHz}$.}
    \label{fig:FMR}
\end{figure}

\textit{Model.}---The Landau-Lifshitz-Gilbert (LLG) equation 
\begin{align}
    \label{eq:LLG}
    \partial_t\M-\frac{\alpha}{M_s} \M\times\partial_t\M =& -\gamma  \M \times \mathbf{H}_{\rm eff} \\
    &+\frac{\tau}{M_s}\M\times(\M\times \mathbf{e}_z),\nonumber
\end{align}
governs the dynamics of the Kittel mode~\footnote{Detailed calculations are provided in the Supplemental Material.}, where $\M$ is the magnetization with constant modulus $M_s$, $\alpha$ is the Gilbert damping, $\gamma$ the gyromagnetic ratio, and $\tau \sim \theta_s g j$ is the strength of STT proportional to the spin Hall angle \(\theta_s\), the mixing conductance \(g\) and the applied current density \(j\).
We adopt an effective field $\mathbf{H}_{\rm eff}=(h_z+\kappa_0 M_z/M_s)\mathbf{e}_z+\mathbf{h}_M^0+\mathbf{h}_\perp(t)$ with external magnetic field $h_z\mathbf{e}_z$, uniaxial crystalline anisotropy constant $\kappa_0$, the demagnetization field $\mathbf{h}_M^0=-N_z M_z\mathbf{e}_z-N_\perp M_\perp\mathbf{e}_\perp$ with demagnetization tensor coefficients $N_z$ and $N_\perp$, and transverse field $\mathbf{h}_\perp(t)=h_\perp(\mathbf{e}_x\cos\omega t+\mathbf{e}_y\sin\omega t)$.
We take the crystalline anisotropy of the film to compensate its shape anisotropy, so that the net perpendicular anisotropy vanishes, $\kappa_0=\kappa+M_s(N_z-N_\perp)$~\cite{Kurebayashi_Barker_Yamazaki_Kushwaha_Stenning_Youel_Hou_Dion_Prestwood_Bauer_etal._2026,Karadza_Wang_Kercher_Noel_Legrand_Schlitz_Gambardella_2026,AnisotropicCompensation}.
The uniform solution of Eq.~\eqref{eq:LLG} is the unit vector
\begin{align}
\label{eq:M0}
    \M_0/M_s=&\sin\theta_0\cos(\omega t-\phi_0)\mathbf{e}_x \nonumber\\
    &+\sin\theta_0\sin(\omega t-\phi_0)\mathbf{e}_y+\cos\theta_0\mathbf{e}_z,
\end{align}
which depends on the cone angle $\theta_0$ and the phase lag $\phi_0$ relative to the drive.
Ferromagnetic resonance (FMR) spectroscopy monitors uniform magnetization dynamics, or the Kittel mode, via microwave absorption power. 
The absorbed power, $p\equiv \mathbf{h}_\perp\cdot\partial_t\mathbf{M}_0$~\cite{BERTOTTI2009153}, is shown in Fig.~\ref{fig:FMR}.
Figure~\ref{fig:FMR}(a) compares the FMR absorption with and without an STT of $\tau=2\tau_c$, where $\tau_c=\alpha\gamma(h_z+\kappa)$ is the torque that compensates Gilbert damping and stabilizes the inverted magnetization.
The inverted response is exactly the negative of the ordinary one~\footnote{The negative absorbed power reflects the negative-energy character of the uniform excitation on top of the inverted magnetization.}.
For sufficiently large $h_\perp$, both curves display hysteresis, the foldover effect~\cite{foldover:Chen_Patton_Srinivasan_Zhang_1989,foldover:Fetisov_Patton_Synogach_1999,foldover:PhysRevB.80.184422}.
To leading order in the drive, the foldover can be mapped onto the driven anharmonic Duffing oscillator with two stable points~\cite{Duf:Moon2014,Duf:PhysRevE.111.064202}.
The foldover observed here arises because the anisotropy allows two dynamically stable solutions for $(\theta_0,\phi_0)$.
Figure~\ref{fig:FMR}(b) adds a complementary observation, where a moderate torque, $\tau<\tau_c$, combined with a transverse drive strongly reduces the foldover threshold.
Therefore, even a moderate STT may induce nonlinear magnetization dynamics with much smaller microwave excitation amplitudes. 
In the following, we assume that the anisotropy is small enough to suppress bistability, which implies that the steady solutions are simple precessions. The spin waves are then the sole source of nonlinearities caused by dipolar and/or exchange interactions.
We define spin waves as the small deviations from the uniform precession, $\M(\mathbf{r},t)=\M_0(t)+\dm(\mathbf{r},t)$ with $\M_0\cdot\dm=0$ and $|\dm|\ll M_s$.
The magnetostatic field to leading order in  \(\dm \) reads $\mathbf{h}_M=\mathbf{h}_M^0+\delta \mathbf{h}_M$ where $\delta\mathbf{h}$ are the fluctuations in the magnetostatic field.
Following Refs.~\cite{BERTOTTI2009193,STTdyn:Bertotti_2009}, the time dependent texture obeys

\begin{widetext}
\begin{align}
\label{eq:pertLLG}
    \partial_t\dm-\frac{\alpha}{M_s} \M_0\times\partial_t\dm =& - \dm \times \left(\gamma\mathbf{H}_{\rm eff}^{(0)}-\frac{\alpha}{M_s} \partial_t\M_0-\frac{\tau}{M_s} \M_0\times \mathbf{e}_z\right)-\M_0\times \left(\gamma \mathbf{H}_{\rm eff}^{(1)}-\frac{\tau}{M_s}\dm\times \mathbf{e}_z\right),
\end{align}
\end{widetext}

\noindent with $\mathbf{H}_{\rm eff}^{(0)}=(h_z+\kappa M_{0,z}/M_s)\mathbf{e}_z+\mathbf{h}_\perp(t)$ and $\mathbf{H}_{\rm eff}^{(1)}=\mu_0\delta \mathbf{h}_M+\kappa \delta m_{z}\mathbf{e}_z/M_s+2A\nabla^2\dm/M_s$, where $\M_0$ is the time dependent uniform magnetization in Eq.~\eqref{eq:M0} and $A$ is the exchange stiffness.
For thin films with magnetostatic boundary conditions, plane waves $\dm(t)e^{i \mathbf{q}\cdot \mathbf{r}}$ in Eq.~\eqref{eq:pertLLG} lead to a linear differential equation, which cannot be solved analytically. Here, the in-plane wave vector $\mathbf{q}$ is defined by its amplitude $q$ and angle $\vartheta_q$ with respect to \(\hat z\).

\begin{figure}[t]
\includegraphics[width=\linewidth]{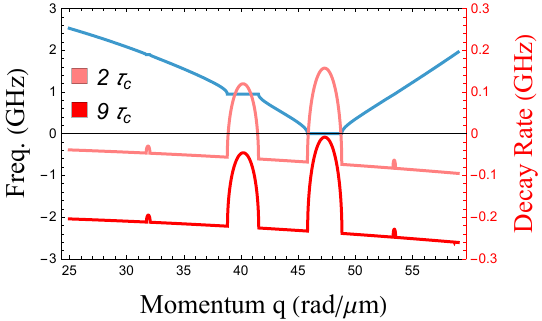}
\caption{\label{fig:StabilityofDisp}%
Antimagnon dispersion relations (blue curves), $\mathrm{Re}[\nu_m(a,b)\omega /2]$, and decay rate (red), $-\eta(t)/t-\mathrm{Im}[\nu_m(a,b)\omega /2]$, in the rotated frame for $\vartheta_q=\pi/4$, shown for $\tau=2\tau_c$ (light red) and $\tau=9\tau_c$ (dark red).
Wherever the red curve exceeds zero, the solution grows exponentially in time, so stability imposes a lower bound on the STT. Here $\tau=9\tau_c$ suffices to stabilize the parametrically driven antimagnons.
The parameters are as in Fig.~\ref{fig:spectra}.}
\end{figure}

\begin{figure}
    \centering
\includegraphics[width=\linewidth]{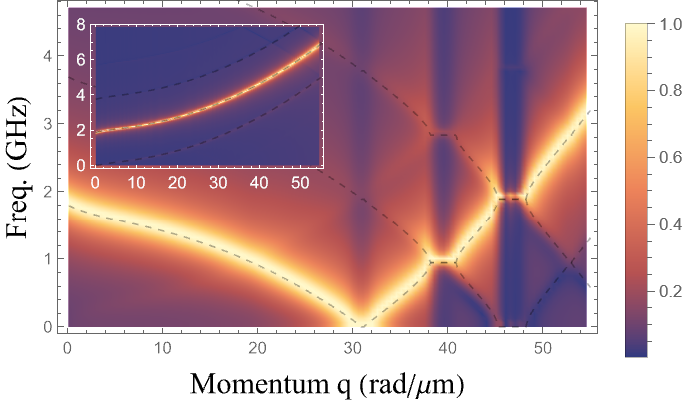}
    \caption{Numerically computed spectral densities of spin waves on top of the magnetization precessing with cone angle $\theta_0=\pi-0.3$, overlaid by the analytical antimagnon dispersion at an equivalent cone angle (dashed lines) for $\gamma h_z=\omega=1.89~\mathrm{GHz}$, $\gamma h_\perp=56~\mathrm{MHz}$, $A=2.8~\mathrm{pJ/m}$, $\kappa=0$, $\omega_M=\gamma\mu_0 M_s=1.11~\mathrm{GHz}$, $d=0.5~\mu$m, $\vartheta_q=\pi/4$, and $\tau=9\tau_c$ as required by Fig.~\ref{fig:StabilityofDisp}.  The inset shows the magnon spectral density for the same parameters except $\tau=0$ and $h_\perp$ tuned so that $\theta_0=0.3$ without significant deviations from parabolicity, whereas the antimagnon spectrum shows frequency-locked plateaus near $\omega/2$ and $\omega$ accompanied by a frequency comb.}
    \label{fig:spectra}
\end{figure}

\begin{figure}[t]
\includegraphics[width=\linewidth]{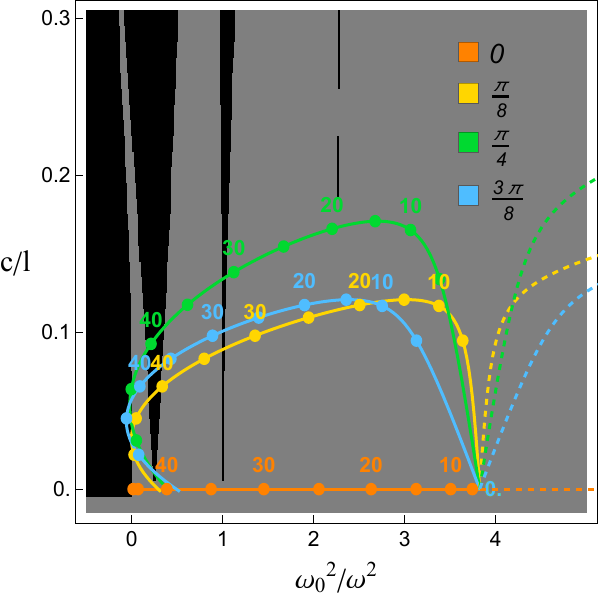}
\caption{\label{fig:Stability_analysis}%
Comparison of the stability of the Kapitza pendulum with that of magnetization dynamics. Dark regions (Mathieu tongues) indicate combinations of the drive amplitude \textit{c} relative to the pendulum length $c/l$ and the squared ratio $\omega_0^2/\omega^2$ of the harmonic and drive frequencies that render the Kapitza pendulum unstable, while the gray regions indicate stable, small-amplitude oscillations. These regions were acquired by numerically evaluating the solutions to the linearized Eq.~\eqref{eq:kapitza}. 
The colored dots illustrate the mapping of the parameters of the Kapitza pendulum on the driven spin waves $((\omega_0/\omega)^2,c/l)\Leftrightarrow(a(q,\vartheta_q),b(q,\vartheta_q))/4$; those connected by solid lines are for antimagnons and dashed lines for magnons for $q\in[0,50]~\mathrm{rad}/\mu$m. Neighboring dots correspond to $\Delta q = 5~\mathrm{rad}/\mu$m. Color encodes the angle $\vartheta_q$ as indicated.
The antimagnon path crosses the low-order tongues, whereas that of the magnons moves in the opposite direction.}
\end{figure}

The expansion of linearized Eq.~\eqref{eq:pertLLG} in $\theta_{0}$ can be simplified  by the transformation $\dm=S(t)\dm'$, where~\cite{BERTOTTI2009193},

\begin{align}
    S(t)=&\left(1-\sin ^2\theta_0 \cos ^2(\omega t)\right)^{-1}\times\nonumber\\
    &\left(
\begin{array}{cc}
 \cos \theta_0 \cos (\omega t) & \sin (\omega t) \\
 -\sin (\omega t) & \cos \theta_0 \cos (\omega t) \\
\end{array}
\right).
\end{align}
squeezes and rotates the magnon amplitudes. We can solve the problem analytically for small deviations of $\theta_0$ from the equilibrium ($\theta_0=0$) and inverted ($\theta_0=\pi$) configurations and microwaves tuned to the resonance $\omega=\gamma (h_z+\kappa \cos\theta_0)$.  In the Supplemental Material (SM) we derive the equation of motion for the primed amplitudes 
\begin{align}\label{eq:dm1}
    \partial_t^2\delta m_1'+\beta(t) \partial_t\delta m_1'+\nu^2(t)\delta m_1'=0,
\end{align}
where $\beta(t)$, $\nu(t)$ are the damping coefficient and precession frequency, respectively.
Close to the inverted state ($\theta_0\rightarrow\pi$), and when $h_\perp=0$, $\nu(t)=\sqrt{\omega_{\rm am}^2(\mathbf{q})-\half\omega_M^2g_q^2\sin^4\vartheta_q}$ is a constant in time and equivalent to the antimagnon dispersion relation with
\begin{equation}
\label{eq:wam}
\omega_{\rm am}(\mathbf{q})=-\frac{2\gamma A q^{2}}{M_s}+\gamma h_z-\frac{1}{2}\omega_M\,g_q\sin^{2}\vartheta_q,
\end{equation}
where $g_q=1-(1-e^{-qd})/qd$ and $\omega_M=\gamma\mu_0 M_s$. The essential difference with the conventional magnon dispersion is the negative sign of the quadratic exchange term.
Equation~\eqref{eq:dm1} may be further simplified for small cone angles by the ansatz $\delta m_1'(t)=f(t)\exp(-\eta(t))$ with $\eta(t)=\half\int\beta(\tilde t) d\tilde t=(\tau-\alpha\,\omega_{\rm am}(q))t$ to
\begin{align}
\label{eq:feom}
    \partial_t^2 f+\Omega^2(t) f=0,
\end{align}
where $\Omega^2(t)=\nu^2(t)-\tfrac{1}{2}\dot\beta(t)-\tfrac{1}{4}\beta^2(t)$. When $\alpha\ll|2 \alpha \tau+3 \omega+2\omega_{\rm am}(q)|$,
\begin{align}
\label{eq:ab}
    \Omega^2(t)\approx \frac{\omega^2}{4}a(q,\vartheta_q)-\frac{\omega^2}{2}b(q,\vartheta_q) \cos \omega t .
\end{align}
For the parameters used in Figs.~\ref{fig:StabilityofDisp}~-~\ref{fig:Stability_analysis}, this condition is satisfied for $q<54~\mathrm{rad}/\mu$m. The coefficients $a$ and $b$ for both antimagnons and magnons are derived in the SM.
The Mathieu equation [Eq.~\eqref{eq:feom}]~\cite{NSW:Bauer1} illustrates that the dipolar field of the uniform mode parametrically drives the spin-wave excitations.

The solution of Eq.~\eqref{eq:feom} is $f(t)=P(a,b,\omega t/2)\exp[i\nu_m(a,b)\omega t/2]$, so $\delta m_1'(t)=P(a,b,\omega t/2)\exp[i\nu_m\omega t/2-\eta(t)]$, where $\nu_m$ is the Mathieu exponent and $P$ is an envelope function that oscillates at the drive frequency.
The exponent contains the dispersion $\mathrm{Re}[\nu_m\omega /2]$ and the decay rate $-\eta(t)/t-\mathrm{Im}[\nu_m\omega /2]$, plotted for antimagnons in Fig.~\ref{fig:StabilityofDisp}.
We observe flat regions in the dispersion in which the decay is reduced and even changes sign. In the latter case, the perturbations are amplified exponentially in time, and our approach breaks down \cite{Chaos:Suhl_Zhang_1986,Chaos:Zhang_Suhl_1985,Chaos:Gibson_Jeffries_1984,Chaos:DeAguiar_Rezende_1986,Chaos:Rezende_deAlcantaraBonfim_deAguiar_1986}.
Stability can be restored by increasing the STT, as shown in  Fig.~\ref{fig:StabilityofDisp} for $\tau=9\tau_c$, which keeps the dispersion nearly unchanged. 

\textit{Enhanced antimagnon nonlinearity.}---The dispersion in the laboratory frame is plotted as dashed lines in Fig.~\ref{fig:spectra} on top of the spectral densities obtained by a numerical solution of Eq.~\eqref{eq:pertLLG}.
Where available, the analytic solutions agree surprisingly well with the simulations, illustrating that the physics is robust up to relatively large precession angles and creating confidence in the numerical results in the regime $\omega\neq\gamma (h_z+\kappa \cos\theta_0)$ presented in the SM.
Figure~\ref{fig:spectra} displays the two central observations: flat regions in the dispersion indicating frequency locking as well as frequency combs.
Note that the antimagnon dispersions grow soft at $q\approx31~\mathrm{rad}/\mu$m.
Here, the STT drive suppresses an instability expected otherwise for zero-frequency magnons~\cite{soft_magnons}.

It is instructive to compare the precessional dynamics at the same cone angle relative to the equilibrium and inverted magnetization.
The time-dependent effects predominantly depend on $\theta_0$, as seen by $b(q,\vartheta_q)$ in the SM. 
Since the large STT needed to stabilize the inverted magnetization enhances the damping, we must apply a larger microwave power to reach a given $\pi-\theta_0$.
At equal drive amplitude and $\tau=2\tau_c$, the antimagnon band strongly deviates from the parabolic dispersion and instabilities may arise at weak microwave drives [see Fig.~\ref{fig:StabilityofDisp}]. 
When the STT is close to but larger than \(\tau_c\), the interactions destabilize the inverted state, indicating that \(\tau_c\) is a lower-bound estimate of the critical torque. 

We visualize the difference between magnons and antimagnons by a mechanical model for Eq.~\eqref{eq:feom}, viz. the ``Kapitza" rigid pendulum driven by an oscillating pivot with equation of motion
\begin{align}
\label{eq:kapitza}
    \partial_t^2\varphi+\left(\omega_0^2+\omega^2\frac{c}{l}\cos\omega t\right) \sin \varphi=0,
\end{align}
where $\varphi \in [0,\pi]$ is the pendulum angle, $\omega_0$ the eigenfrequency of the pendulum at small angles without drive, $l$ its length, while $c$ and $\omega$ are the amplitude and frequency of the pivot \cite{Kapitza_1951,Butikov_2001}.
Whether its solutions grow or decay depends only on the dimensionless parameters $\omega_0/\omega$ and $c/l$.
Figure~\ref{fig:Stability_analysis} maps the solutions of Eq.~\eqref{eq:kapitza} for $\varphi\ll1$ to either stable (gray) or unstable (black) regions.
The black areas are known as Mathieu tongues.
Each tongue, labeled $n=0,1,2,\dots$ from the left, marks a resonance of the drive with the natural frequency. Inside the $n$-th tongue, the pendulum oscillates at $n\omega/2$ even when $\omega_0$ is somewhat detuned.
We observed this frequency locking via the flat regions of the dispersion in Figs.~\ref{fig:StabilityofDisp}~-~\ref{fig:spectra}.

Mapping the coefficients of Eq.~\eqref{eq:ab} onto the parameters of the Kapitza pendulum via $((\omega_0/\omega)^2,c/l)\Leftrightarrow(a(q,\vartheta_q),b(q,\vartheta_q))/4$ allows a direct comparison with the spin-wave dynamics in the ``phase diagram" of Fig.~\ref{fig:Stability_analysis}.
The antimagnon curves in the $((\omega_0/\omega)^2,c/l)$ space (dots connected by solid curves) cross the low-order tongues for various  $q$ and $\vartheta_q$, whereas the conventional magnon trajectories (dots connected by dashed curves) avoid them. Indeed, magnon frequencies increase with $q$, so $(\omega_0/\omega)^2$ cannot access the higher harmonic resonances, while the antimagnon frequencies decrease with wave number and move in the opposite direction in the phase diagram to cross the lowest resonance. We may conclude that, keeping everything else the same, the parametric pumping of antimagnons requires much less power than that of conventional magnons.

\textit{Frequency combs and level attraction.}---The frequency combs in Fig.~\ref{fig:spectra} are a direct consequence of the Floquet equation [Eq.~\eqref{eq:feom}] that is invariant under time translation $t\to t+2\pi/\omega$. The envelope $P(a,b,\omega t/2)$ in the $\delta m^\prime$ solution is periodic, generating replicas shifted by integer multiples of $\omega$ in the spectral density.
These replicas cross, and in the inverted state each crossing pairs a positive- with a negative-energy branch, i.e., magnons and antimagnons on top of the inverted magnetization.
This can be captured by two coupled harmonic oscillators
\begin{align}
\label{eq:H}
    \hat H=\hbar \sum_{q}\Big\{\omega_{\rm m}(q)\hat b^\dagger_q \hat b_q-\omega_{\rm am}(q)\hat a^\dagger_q \hat a_q+\lambda\hat a^\dagger_q \hat b^\dagger_{-q}+\lambda^* \hat a_q\hat b_{-q}\Big\},
\end{align}
where $\hat a_q$ and $\hat b_q$ are antimagnon and magnon operators with frequency $\omega_{\rm am}(q)$ and $\omega_{\rm m}(q)$. The coupling looks like a parametric drive with coupling strength $\lambda$ that is of the same order as $b(q,\vartheta_q)$. The eigenvalues read
\begin{align}
\label{eq:omegapm}
    \omega_{\pm}=\frac{1}{2}\left(\omega_{\rm m}+\omega_{\rm am}\pm \sqrt{(\omega_{\rm m}-\omega_{\rm am})^2-4|\lambda|^2}\right).
\end{align}
Since $\omega_{\rm am}(q)>0$ for $q<\sqrt{M_sh_z/2A}$, the square root turns imaginary when $(\omega_{\rm m}-\omega_{\rm am})^2<4|\lambda|^2$, i.e., near the crossing.
The ``level attraction" caused by the coupling of positive- and negative-frequency modes is bounded by two exceptional points in parameter space~\cite{LevelAttraction:PhysRevA.98.023841}.
This analysis applies to the flat regions in the dispersion relation, which are crossings between positive and negative-energy branches.
The imaginary component acquired by the frequencies according to Eq.~\eqref{eq:omegapm} at these crossings signals a parametric instability and explains why the flat regions are linked to gain in Fig.~\ref{fig:StabilityofDisp}.
As noted above, this gain is offset by increasing the STT.
Tuned just above the stability threshold, the parametric gain and the STT-induced damping nearly cancel, so the net decay rate approaches zero and the spectral lines in Fig.~\ref{fig:spectra} narrow accordingly.
The comb teeth then combine a high quality factor with the rigid spacing $\omega$ imposed by the drive, which is what makes them useful for frequency conversion.
The coupling in Eq.~\eqref{eq:H} moreover creates and annihilates magnon and antimagnon pairs, so the flat regions within the dispersion relation are populated by two-mode-squeezed, entangled pairs, which may be a resource for quantum magnonic applications.

\textit{Conclusion.}---The non-linearities of antimagnons on top of a driven inverted magnetization are much more pronounced than those of magnons because of their proximity to parametric resonances driven by the dipolar field from the uniform precession around the inverted magnetization. This leads to 
(i) frequency locking: an input at $\omega$ generates a large population of antimagnons at exactly $\omega/2$ or $\omega$, depending on the probed wave vector.
(ii) pronounced frequency combs that can enable precise frequency conversion in magnonic devices.
These predictions can be experimentally tested by Brillouin light scattering \cite{AntimagnonEXP}. While the weak exchange in yttrium iron garnets limits access to the small wave vector region of Fig.~\ref{fig:spectra}, Brillouin light scattering may provide more access for other materials with strong exchange or by using stronger driving at lower frequencies, as shown in the SM.
(iii) A weak microwave input can destabilize an inverted magnetization that therefore may act as a trigger or sensor detectable by microwave emission or absorption \cite{Chaos:Gibson_Jeffries_1984}.
Our analysis holds only for weak perturbations on top of the macrospin dynamics. We expect a rich hierarchy of higher-order nonlinearities, including Suhl instabilities, domain-like ordering, and chaos.
(iv) The level attraction should lead to exceptional-point dynamics and anomalous magnon-antimagnon correlations, which qualify inverted magnets as a platform for quantum control of magnons and antimagnons.

\begin{acknowledgments}
A.~L.~B. thanks Rob den Teuling for fruitful discussions.
Claude Opus 5 has been used for text editing purposes.
A.~L.~B. and R.~A.~D. acknowledge support from ``Black holes on a chip'' (project number OCENW.KLEIN.502) and ``Fluid Spintronics'' (project number VI.C.182.069), both financed by the Dutch Research Council (NWO).
G.~E.~W.~B. acknowledges support from JSPS KAKENHI Grants No. 22H04965 and No. 24H02231.
M.~E. acknowledges support from JST ASPIRE (JPMJAP26A1).
\end{acknowledgments}

\bibliographystyle{apsrev4-2}
\bibliography{bib}

\end{document}